\documentstyle[11pt]{article}
\newcommand{\be}{\begin{equation}}
\newcommand{\ee}{\end{equation}}
\newcommand{\bea}{\begin{eqnarray}}
\newcommand{\eea}{\end{eqnarray}}
\newcommand{\f}{\frac}

\newcommand{\te}{\theta}

\newcommand{\vs}[1]{\vspace{#1 mm}}
\newcommand{\hs}[1]{\hspace{#1 mm}}

\begin{document}
\hsize=5.6truein
\vsize=9truein
\hoffset=-.3in
\textheight=8.5truein
\voffset=-.8truein
\baselineskip=.60cm
\rightline{hep-th 9911183}

\vs{20}

\centerline{\large\bf Scalar Absorption by Noncommutative  D3-branes}
\vs{10}
\centerline{Ali Kaya\footnote{e-mail: ali@rainbow.tamu.edu}} 
\vs{5}
\centerline{Center for Theoretical Physics, Texas A\& M University,}
\centerline{College Station, Texas 77843, USA.}
\vs{15}
\begin{abstract}

The classical cross section for low energy absorption of the RR-scalar by
a stack of noncommutative D3-branes in the large NS B-field limit is
calculated. In the spirit of AdS/CFT correspondence, this cross section is
related to two point function of a certain operator in noncommutative
Yang-Mills theory. Compared at the same gauge coupling, 
the result agrees with that of obtained from ordinary D3-branes. This is 
consistent with the expectation that ordinary and noncommutative
Yang-Mills theories are equivalent below the noncommutativity scale, but
it is a nontrivial prediction above this scale.

\end{abstract}
\vs{50}
\pagebreak

Theories on noncommutative  spaces  arise naturally in string-M
theory. In \cite{pmh}, the system of parallel D-branes has been
reinterpreted as a quantum space \cite{cones}. DLCQ of M-theory with
nonzero background three-form field along the null compactified direction
has been argued to have matrix theory description of the gauge theory on
noncommutative torus \cite{cds}. It has been shown in \cite{dh} that the
D-brane world volume theories are noncommutative at a certain limit of
the compactification moduli. Applying Dirac's constrained  quantization 
method to open strings in the presence of NS two-form (B) field, the
spacetime coordinates of open string end points have been shown not to
commute \cite{ch1}\cite{arda}. Recently, by a direct string theoretic
analysis, it has been shown in \cite{sw} that noncommutativity in the
effective action is natural in the presence of constant background
B-field.\\

A particularly interesting example of all these is the  noncommutative
Yang-Mills theory (NCYM) in four dimensions. The spectrum of  IIB theory 
in the presence of D3-branes on constant B-field consists of open and
closed string excitations coupled together. However, it is possible to take
a low energy limit and scale some parameters such that the closed string
modes decouple \cite{sw}. The resulting theory of open string modes turn
out to be the NCYM. In \cite{hi}\cite{mr}, a supergravity background has
been proposed to be dual to this system. This background can be obtained
by first constructing a solution which has nontrivial  B-field dependence
and then taking the decoupling limit of \cite{sw}. Closed string two point
scattering amplitudes in the presence of  B-field have been calculated and
shown to be consistent with the ones encoded in the gravity solution
\cite{sh}.  In \cite{lw}, the solution has been shown to have holographic
features. Some properties of the NCYM have also been studied using
its gravity dual \cite{mr}\cite{hh}\cite{co}.\\

In this paper we will consider the process of classical low energy 
absorption of the RR-scalar by noncommutative D3-branes in the large
B-field limit.
In the spirit of AdS/CFT correspondence, the cross section
calculated from the gravity side is related to discontinuity of two point
function of a certain operator in the dual NCYM \cite{gk3}. This operator
can be deduced from the coupling of the scalar to the  D-brane world
volume effective action \cite{dt}.  The fluctuations turn out to be
non-minimally coupled  to background geometry. However, the coupling of
RR-scalar to the effective world volume theory is
relatively simple which may help one to identify the
operator in NCYM. In calculating the absorption cross section, we will
work with the D3-brane solution of \cite{mr} before taking the decoupling
limit. As discussed in the context  of ordinary D3-branes in \cite{gk4},
the decoupling limit identifies the so called throat region with a
certain limit \footnote{The conformal limit in the case of ordinary  
D3-branes.} of the world volume theory. On the other hand, a
large B-field is also encountered in the decoupling limit considered in
\cite{sw} \footnote{In taking the decoupling limit one keeps B fixed and 
let $\alpha '$ go to zero which corresponds to a very large B-field for
finite (but small) $\alpha '$.}, which means that the dual theory is
"close" to its throat limit.
For previous work on cross section calculations on black-brane
backgrounds see, for instance, \cite{das2}-\cite{gh}.\\

The type IIB supergravity action in the string frame can be written as
\bea \label{action}
S_{IIB}=\f{1}{2\kappa_{10}^{2}}\int dx^{10} \sqrt{-g} ( e^{-2\phi} [R
+ 4 (\partial \phi )^{2} -\f{3}{4} (\partial B_{2})^{2}] \nonumber\\ 
-\f{1}{2} (\partial C)^{2} - \f{3}{4} (\partial C_{2} - C \partial
B_{2})^{2} -\f{5}{6} F(C_{4})^{2} -\f{\epsilon_{10}}{48} C_{4}\partial
C_{2} \partial
B_{2} +... ) 
\eea
where $\partial B_{2} = \partial_{[\mu} B_{\nu\lambda]}$
etc., $F(C_{4}) = \partial C_{4} + 3/4 (B_{2} \partial C_{2} - C_{2}
\partial B_{2} ) $, and the self duality of $F(C_{4})$ is imposed at the
level of field equations. Specifically $B_{2}$ denotes the NS two-form
field and $C$ is the RR-scalar.  The solution corresponding to
noncommutative D3-branes is given by \cite{mr}, 
\be \label{geo}
ds^{2}= f ^{-1/2} [-dt^{2} + dx_{1}^{2} + h (dx_{3}^{2} + dx_{4}^{2})]
+ f^{1/2} [dr^{2} + r^{2} d\Omega_{5}^{2}], 
\ee 
\be 
B_{23}=\f{\sin \te}{\cos \te}f^{-1}h, \hs{5} e^{2\phi}=g_{s}^{2}h, \hs{5}
\partial (C_{2})_{01r}=\f{1}{g_{s}} \sin \te \partial_{r} f^{-1},\hs{3} 
\ee
\be
F(C_{4})_{0123r}=\f{1}{g_{s}} \cos\te h \partial f^{-1},\hs{54}
\ee 
where $g_{s}$ is the asymptotic value of the string coupling constant, and
the
functions $f$ and $h$ are given by
\be
f=1+\f{R^{4}}{r^{4}}, \hs{10} h= \f{f}{\sin^{2}\te  + \cos^{2} \te f}.
\ee
The solution is asymptotically flat and there is a horizon at $r=0$. The
near horizon geometry is $AdS_{5}\times S_{5}$. The mass per unit volume
of (\ref{geo}) in string frame can be calculated to be
\be\label{mass}
M=\f{2\pi^{3}R^{4}}{\kappa_{10}^{2}g_{s}^{2}},
\ee
which is remarkably independent of $\te$.  Note that the asymptotic value
of B-field is $\tan \te$.\\ 

From (\ref{action}), the fluctuations of the RR-scalar on this background
obeys 
\be\label{7}
\nabla^{2} C = \f{3}{2} (\partial B_{2})^{2} C,
\ee
and thus couple to the background geometry non-minimally. We note
that the contraction $(\partial B_{2})(\partial C_{2})$ is zero.
Respecting the translational invariance,
we will assume that $C$ does not depend on the spatial coordinates of
D3-branes. Separating the time dependence and considering a spherically
symmetric fluctuation (s-wave), which is supposed to give the dominant
contribution to cross section, we write
\be\label{88}
C= e^{-iwt} \phi (r).
\ee
Following from (\ref{7}), $\phi$ obeys
\be\label{ana}
(hr^{5})^{-1}\f{d}{dr}(hr^{5}\f{d}{dr}\phi) + \omega^{2}f\phi -
\f{16\sin^{2}\te \cos^{2}\te R^{8}}{r^{10}} f^{-3}h^{2}\phi =0.
\ee
Since this equation does not appear to be analytically solvable,
following
previous work, we try to find an approximate solution by matching three
different regions dictated by the structure of the functions $f$ and $h$.
Low energy scattering is characterized  by $\omega\sqrt{\alpha '} \ll 1$
and the $\alpha '$
corrections to  background is suppressed when $\sqrt{\alpha '}/R \ll 1$. 
Consistent with these two restrictions, we will consider the double
scaling limit of \cite{ik} and assume $\omega R \ll 1$. Large B-field
corresponds to $\cos\te \ll 1$ and we will further analyze the case
where $\cos\te \sim \omega R$. 
\subsection*{Region I: $r\gg R$}

In this region $f\sim h \sim 1$. Defining $\rho =\omega r$ and
$\phi = \rho^{-5/2} \psi$, (\ref{ana}) simplifies as
\be\label{rg1}
\left(\f{d^{2}}{d\rho^{2}}  - \f{15}{4\rho^{2}}  + 1 -
\f{16\sin^{2}\te\cos^{2}\te (\omega R)^{8}}{\rho^{10}}\right) \psi =0.
\ee
Since $\rho \gg \omega R$, the last term in this equation is negligible
compared to the second one. Ignoring this term, (\ref{rg1}) can be solved
in terms of Bessel and  Neumann functions which in turn gives
\be \label{rgn1}
\phi = a_{1} \hs{2}\rho^{-2} J_{2}(\rho ) + a_{2}\hs{2} \rho^{-2} N_{2}
(\rho ),
\ee
where $a_{1}$ and $a_{2}$ are constants.
\subsection*{Region II: $R\gg r \gg R\sqrt{\cos\te}$}

In this region $f$ and $h$ can be approximated as $f \sim h \sim
R^{4}/r^{4}$. Using this form and defining $\phi = \rho^{-1/2}\chi$,
(\ref{ana}) becomes
\be\label{rg2}
\left( \f{d^{2}}{d\rho^{2}}  + \f{1}{4\rho^{2}}  + \f{(\omega
R)^{4}}{\rho^{4}}  - \f{16\sin^{2}\te\cos^{2}\te (\omega
R)^{4}}{\rho^{6}} \right) \chi =0.
\ee
In region II, $\omega R \gg\rho \gg \omega R \sqrt{\cos\te}$ and in that
interval;  the third term can  be ignored compared to the fourth one
(with the assumption $\cos\te \sim \omega R$), and the fourth term is
always very small with respect to the second one. Therefore, $\chi$  
approximately obeys
\be
\f{d^{2}}{d\rho^{2}}\chi + \f{1}{4\rho^{2}} \chi = 0.
\ee
Two solutions of this equation are $\rho^{1/2}$ and $\rho^{1/2}\ln
\rho$, and thus
\be\label{rgn2}
\phi = b_{1} + b_{2}\hs{1} \ln\rho,
\ee
where $b_{1}$ and $b_{2}$ are constants.
\subsection*{Region III: $R\sqrt{\cos\te}\gg r$}

In this region $f\sim R^{4}/r^{4}$ and $h\sim 1/\cos^{2}\te$.
Defining $z= \omega R \sqrt{\cos\te}/ \rho$ and $\phi = z^{3/2}Z$,
equation (\ref{ana}) can be approximated to,
\be
\left( \f{d^{2}}{dz^{2}}  -\f{15}{4z^{2}}  + \f{(\omega R)^{2}}{\cos\te}
- 16 \f{\sin^{2}\te}{z^{6}} \right) Z =0.
\ee
In region III, $z\gg 1$ and thus the last term can be ignored compared
to the second one. Dropping this term, two solutions of $Z$ can be
found to be  $z^{1/2} J_{2} (z\omega R/ \sqrt{\cos\te})$ and 
$z^{1/2} N_{2} (z\omega R/ \sqrt{\cos\te})$. This gives
\be\label{rgn3}
\phi = c_{1} \hs{2}\rho^{-2} J_{2}\left(\f{(\omega R)^{2}}{\rho}\right) +
c_{2} \hs{2} \rho^{-2} N_{2}\left(\f{(\omega R)^{2}}{\rho}\right), 
\ee
where $c_{1}$ and $c_{2}$ are constants.
\subsection*{Matching the solutions:} 
In matching the solutions in different regions, we will use the small
argument expansion of the Bessel and Neumann functions
\bea\label{*}
J_{2}(x)&\sim& \f{x^{2}}{8},\nonumber\\
N_{2}(x)&\sim& \f{-4}{\pi x^{2}}(1+\f{x^{2}}{4}) + \f{1}{4\pi} x^{2}(\ln x
+c),
\eea
where $c$ is a constant.  Close to the horizon, we want only an ingoing
wave which implies 
\be 
c_{1}=- i \hs{1}c_{2}.
\ee
The overall normalization of $\phi$ can be fixed by
imposing $c_{1}=i(\omega R)^{4}$. To be able to match the solution
(\ref{rgn3}) to region II, we consider its behavior when $\rho\sim \omega
R\sqrt{\cos\te}$. Assuming $\cos\te\sim \omega R$,  the arguments of the
Bessel and Neumann functions in (\ref{rgn3}) are small in this
range and thus we can use the expansion (\ref{*}). We see that there is no
term in this expansion to be matched by $\ln \rho$ of region II, therefore,
we should set $b_{2}=0$. The dominant contribution of the rest of the
terms when $\rho\sim \omega R\sqrt{\cos\te}$ is the  constant  $4/\pi$,
which fixes
$b_{1}$ as
\be 
b_{1}=\f{4}{\pi}.
\ee 

To match (\ref{rgn1}) to region II,  we will consider its behavior when
$\rho\sim \omega R$. The arguments of
Bessel and Neumann functions in  (\ref{rgn1}) are also small in this
range. The leading contributions of their expansions are $a_{1}/8$ and
$(-4a_{2})/(\pi\rho^{4})$, respectively. To be able to match these to
(\ref{rgn2}), one should set 
\be 
a_{2}=0,\hs{5} a_{1}=\f{32}{\pi}.
\ee
Combining these, we obtain the following functions in three regions 
\bea
\phi_{I} & =& \f{32}{\pi} \rho^{-2} J_{2}(\rho ),\nonumber\\
\vs{2}
\phi_{II}& =& \f{4}{\pi}, \nonumber\\
\phi_{III}&=&  i(\omega R)^{4} \rho^{-2}\left[J_{2}\left(\f{(\omega
R)^{2}}{\rho}\right)
+i N_{2}\left(\f{(\omega R)^{2}}{\rho}\right)\right] ,  
\label{21} \eea
which smoothly overlaps and give an approximate solution to
(\ref{ana})\footnote{To ensure that $\phi$ can be differentiated twice,
one has to let $b_{2}=O(\omega R)$ and $a_{2}=O((\omega R)^{5})$, instead
of setting them to zero. To zeroth order in $\omega R$, this modification
does not change the main result (\ref{cr}).}. 
To calculate the cross section one should compare the incoming
flux at the horizon with the incoming flux at the infinity. At this stage,
we recognize that the
same functions appear in \cite{ik} in the solutions of the massless wave
equation on ordinary D3-brane background. The cross section
corresponding to the solution (\ref{21}) can be read from \cite{ik} to be
\be\label{cr}
\sigma_{abs} = \f{\pi^{4}}{8} \omega^{3} R^{8}.
\ee
We now try to rewrite  $\sigma_{abs}$ in terms of the gauge coupling
constant $\tilde{g}_{YM}$ of NCYM.\\

The solution (\ref{geo}) can be shown to preserve 1/2 supersymmetries of
the theory. This can easily be seen by nothing that (\ref{geo}) is related
to ordinary D3-brane solution by a chain of T-duality transformations
(namely, first a T-duality along $x_{3}$, then a rotation by an angle
$\te$ along the $x_{2}-x_{3}$ plane, and then another T-duality along
$x_{3}$) and
T-duality respects  supersymmetry when the Killing spinor is independent
of the direction of the duality \cite{berg}. From a world-sheet point of
view, one can also see that the parallel D3-branes on constant,
invertible,   B-field backgrounds also preserve 1/2 supersymmetries since
boundary conditions identify the left moving supercurrents with 
the right moving ones. This is consistent with the identification of  this
configuration with the gravity solution. Due to this BPS property,
the  noncommutative D3-brane tension, when calculated
from an effective action point of view, should be equal to the mass per
unit volume (\ref{mass}). We will now carry out this effective field
theory calculation to fix the value of $R$.\\

The Dirac-Born-Infeld  action corresponding to a noncommutative
D3-brane can be written as 
\be
S_{DBI}= T_{3}\int d^{4}\sigma \sqrt{-\det(\hat{g} + \hat{B}}),
\ee
where $\sigma^{i}$ are coordinates on the D3 brane,  $T_{3}$ is the
ordinary D3-brane tension when $B=0$, $\hat{g}$ and 
$\hat{B}$ are pull-backs of flat Minkowski metric $\eta_{\mu\nu}$ and
constant B-field $B_{\mu\nu}$, respectively,
\bea
\hat{g}_{ij}&=&\partial_{i}X^{\mu}\partial_{j}X^{\nu}\eta_{\mu\nu},  \\
\hat{B}_{ij}&=&\partial_{i}X^{\mu}\partial_{j}X^{\nu}B_{\mu\nu}.
\eea
From $S_{DBI}$, one can calculate the conjugate momentum densities to
coordinate fields $X^{\mu}$ as 
\be\label{mom}
P_{\mu}=\f{\delta S_{DBI}}{\delta \partial_{\tau}X^{\mu}},
\ee
where $\tau$ is the world volume time coordinate. \\

The tension $\tilde{T}_{3}$ of a noncommutative D3-brane can be defined as
the energy density corresponding to a flat, non-exited brane. This can be
described in a physical gauge by $X^{i}=\sigma^{i}$, $X^{\alpha}=const$.
For such a brane, (\ref{mom}) gives
\be
P_{\mu}=T_{3} \sqrt{ - \det(\eta_{ij} + B_{ij})}\hs{2} \delta_{\mu}^{0},
\ee
which corresponds to the momentum density of a massive object with zero
velocity. The energy per unit volume of such an object can be read from 
$P_{0}$ component of the momentum. This gives the tension for
a noncommutative D3-brane as
\be\label{ten}
\tilde{T}_{3}= T_{3} \sqrt{ - \det(\eta_{ij} + B_{ij})}. 
\ee
Note that the ordinary D3-brane tension and the 10-dimensional
gravitational coupling constant are 
\be\label{ordten}
T_{3}= \f{1}{ (2 \pi )^{3} \alpha '^{2} g_{s}}, \hs{7} 
2\kappa_{10}^{2} = (2\pi )^{7} \alpha '^{4}.
\ee\\

In the solution (\ref{geo}), B-field is a rank 2 matrix. Using
(\ref{ten}) and (\ref{ordten}) we obtain,
\be
\tilde{T}_{3}= \f{1}{ (2 \pi )^{3} \alpha '^{2} g_{s} \cos\te}.
\ee
The total energy of $N$-coincident D3-branes is given by
$N\tilde{T_{3}}$ which should be equal to (\ref{mass}). This gives  the
parameter $R$ as\footnote{ Exactly the same expression for $R$ is given in
\cite{mr}.}
 \be\label{R}
R^{4}=\f{4\pi \alpha '^{2} g_{s} N}{\cos \te}.
\ee
On the other hand the gauge coupling constant $\tilde{g}_{YM}$ of NCYM, 
can be read from \cite{sw} to be 
\bea
\tilde{g}_{YM}^{2}&=& 2\pi g_{s} \sqrt{-\det(\eta_{ij} +
B_{ij})},\nonumber \\
                  &=&\f{2\pi g_{s}}{\cos\te}. 
\label{trtr}
\eea
Note that for $\te =0$, this gives the well known relation between the
ordinary Yang-Mills and string coupling constants. 
Combining (\ref{trtr}) with (\ref{R}), the cross section (\ref{cr}) can be
rewritten  in terms of $\tilde{g}_{YM}$ as
\be\label{result}
\sigma_{abs} = \f{\pi^{4}}{2} \omega^{3} \alpha '^{4} N^{2}
\tilde{g}_{YM}^{4}.
\ee
Remarkably, all $\te$ dependence is hidden in $\tilde{g}_{YM}$. The
classical cross section for low energy absorption of RR-scalar by ordinary
D3-branes (which is identical to massless scalar absorption) has been
calculated in \cite{ik} and the result is (\ref{result}) in which
$\tilde{g}_{YM}$ is replaced with the gauge coupling $g_{YM}$ of ordinary
Yang-Mills theory. Comparing NCYM with the ordinary Yang-Mills at the same
coupling 
\be
\tilde{g}_{YM} = g_{YM},
\ee
the result of \cite{ik} exactly agrees with (\ref{result}).  \\

As previously noted, (\ref{result}) is related to discontinuity of the cut
in the two point function of a certain operator in NCYM. The discontinuity
in momentum space is evaluated at $p^{2}=\omega^{2}$. Since
$\omega\sqrt{\alpha '} \ll 1$, (\ref{result}) is valid below the string
scale. On the other hand, since the noncommutativity (NC) scale is roughly  
given by $\sqrt{\alpha ' \cos\te}$ (see, for instance, \cite{sw}) and
$\cos\te \ll 1$, NC scale is also below the string scale. Below the NC
scale, ordinary Yang-Mills theory is a good approximation to NCYM, and the 
equivalence of cross sections is consistent with this fact. Between NC and
string scales, (\ref{result}) is a nontrivial prediction for NCYM. We note
that, (\ref{result}) is the first term in an expansion in the parameter
$(\omega R)$, which in NCYM corresponds to perturbing field theory by
higher and higher dimensional operators and a loop expansion.\\ 

The corresponding operators in dual theories  can be deduced from the
coupling of the scalars at hand to the effective world volume theories.
For the ordinary Yang-Mills theory, leading order coupling of RR-scalar to
the world volume is
\be\label{co}
\epsilon^{ijkl}C \hs{1}{\rm Tr} \hs{1} F_{ij}\hs{1} F_{kl}.
\ee
The classical cross section for the absorption of RR-scalar by ordinary
D3-branes has been shown to agree with a tree level world
volume calculation which involves the above coupling \cite{gk2}. This
indicates (with a non-renormalization theorem) that the leading term of
the corresponding operator in the dual theory is 
$\epsilon^{ijkl} \hs{1} {\rm Tr }\hs{1}F_{ij}\hs{1} F_{kl}$.\\

Naturally, one may try to identify the coupling of RR-scalar to a
noncommutative D3-brane world volume.  As discussed in \cite{sw}, it is 
very convenient to write the effective action in terms of
the noncommutative gauge fields and open string parameters. When expressed
in these variables, the effective action in the presence of B-field can be
deduced form the ordinary one.  From (\ref{co}), the coupling of RR-scalar
to the noncommutative D3-brane world volume can be written as 
\be
\epsilon^{ijkl} C* {\rm Tr} \hat{F}_{ij}* \hat{F}_{kl},
\ee
where $\hat{F}$ is the noncommutative gauge field strength,  
$\epsilon$-tensor and raising of indices refer to open string
metric and $*$-product is defined by
\be
(f*g)(x) = e^{\f{1}{2} \te^{ij}
\f{\partial}{\chi^{i}}\f{\partial}{\eta^{j}}}
f (x+\chi ) g(x+\eta ) |_{\chi = \eta = 0}.
\ee
The open string metric and noncommutativity parameter $\te^{ij}$ is fixed
in terms of the background Minkowski metric and B-field \cite{sw}.
This indicates that the leading order term of the operator in the NCYM is
$\epsilon^{ijkl} \hs{1}{\rm Tr}\hs{1} \hat{F}_{ij}* \hat{F}_{kl}$.
However, this term is {\it not} gauge invariant unless it is integrated
over noncommuting directions.\\

Although it brakes the translational invariance, it is interesting to
consider the effects of adding non-zero momentum along the noncommuting
directions $x_{2}$ and $x_{3}$. For this, one modifies (\ref{88}) as
\be
C= e^{-iwt} e^{i\vec{k}.\vec{x}} \phi (r),
\ee
where $\vec{k}.\vec{x}=k_{2}x_{2}+k_{3}x_{3}$. Due to this modification,
(\ref{ana}) picks up an extra term to the left hand side which is 
\be
-k^{2}fh^{-1}\phi,
\ee
where $k^{2}=k_{2}^{2}+k_{3}^{2}$. It is easy to see that perturbative
scattering requires $\omega > k$. 
In regions II and III, this term can be neglected compared
to another term in (\ref{ana}), namely $\omega^{2}f\phi$, since $h\gg 1$
in these regions and $\omega >k$. On the other hand, in region I,
$\phi_{I}$ is modified
as
\be
\phi_{I} = \f{32}{\pi} (s\rho )^{-2} J_{2}(s\rho ),
\ee
where the parameter $s$ is 
\be
s= \sqrt{ 1 - \f{k^{2}}{\omega^{2}}}.
\ee
The flux at infinity calculated from the modified $\phi_{I}$ is changed
by a factor of $1/s^{4}$ and thus 
\be
\sigma_{abs}(k)= s^{4} \sigma_{abs}(0).
\ee 
For ordinary D3-branes, this factor can be calculated to be $s^{8}$
which seems to imply a disagreement for dual theories. However, we note
that it is difficult to give a world volume interpretation to cross
sections when $s\not = 0 $. For instance, following \cite{ik}, the factor
of $s^{8}$ cannot be obtained by a world volume scattering calculation for
ordinary D3-branes. Furthermore, due to the fact that the corresponding
operator in dual Yang-Mills theory is also a scalar operator, the two
point function of this operator can only depend on $p^{2}$ and 
the discontinuity in the complex plane can only depend on
$\omega^{2}$ (or may be with a slight modification on
($\omega^{2}-k^{2}$)) which is not consistent with the
factor $s^{8}$.  We believe that the broken translational invariance ruins
the connection between cross sections and the two point functions, and
is responsible for these disagreements. Finally, it is also 
worth to mention that the wave equation for minimally coupled
scalars can be solved exactly in Einstein frame and the cross
sections agree to all orders when $k=0$ \cite{mkl}. \\

One may try to repeat same calculations for noncommutative M5-branes using
the solution given in \cite{mr}. For ordinary M5-branes, the
traceless metric perturbations polarized along the 5-branes have been
shown to obey minimally coupled scalar equations \cite{gk2}. One can
easily show that this is also true for noncommutative M5-branes. 
Furthermore, the massless scalar equation on noncommutative M5-branes
turns out to be the same with the one on the ordinary M5 branes.
Therefore, the classical cross sections corresponding to absorption of
traceless metric perturbations polarized along the 5-branes are identical
for both type of branes. \\

{\it Note added for the hep-th version:} After the submission of the
present paper to the net, we received
\cite{mkl} and \cite{mkl2}, which also consider scalar absorption by
noncommutative D3-branes. In \cite{mkl}, the authors claim that to see
effects of noncommutativity one should consider waves propagating along
noncommuting directions on the brane. However, as discussed above,
interpretation of this from a world volume theory point of view is not
clear. In \cite{mkl2}, the authors claim that in the $B\to \infty$,
i.e. $\te\to \pi /2$, limit the noncommutativity effects are turned on,
and in this limit the RR-scalar is nonpropagating.
But, as shown in \cite{sw}, even in the
presence of constant and finite B-field, noncommutativity in the effective
field theory is inevitable. On the other hand, taking $B\to \infty$ limit
is a delicate issue. For instance, to make contact with string theory, the
parameter $R$ should be fixed as in (\ref{R}) and thus diverges at $\te =
\pi /2$. To avoid such infinities, one should  scale other parameters
in a suitable way and this gives the geometry found in \cite{mr}.
Therefore, contrary what is claimed in \cite{mkl2}, the NS 3-form and the
RR 5-form field strengths are not zero at $\te =\pi /2$ . Beside that,
even for the solution in which NS 3-form and RR 5-form field
strengths are zero, one can still introduce fluctuations of these
fields. In either case, it can be shown that RR-scalar is not necessarily
nonpropagating when $\te =\pi /2$.  

\subsection*{Acknowledgements}
I would like to thank S. De\~{g}er for reading the manuscript.

\end{document}